\documentstyle[11pt,newpasp,twoside,epsf]{article}
\def\arg#1{{\it#1\/}}

\def\edcomment#1{\iffalse\marginpar{\raggedright\sl#1\/}\else\relax\fi}
\reversemarginpar

\begin{document}
\title{Photometry and Spectroscopy of Eclipsing Binaries 
in $\omega$~Centauri}
 \author{Janusz Kaluzny}
\affil{Copernicus Astronomical Center, Bartycka 18, 00-716 Warsaw, Poland}
\author{Ian Thompson and Wojtek Krzeminski}
\affil{Carnegie Observatories, 813 Santa Barbara Street, Pasadena, 
CA 91101-1292}
\author{Arkadiusz Olech, Wojtek Pych and Barbara Mochejska}
\affil{Copernicus Astronomical Center, Bartycka 18, 00-716 Warsaw, Poland}

\begin{abstract}
Detached eclipsing double line spectroscopic binaries are accurate
distance indicators.  They can be used for the direct determination of
distances to globular clusters in which such systems can be
identified.  The results of major photometric surveys aimed at
identification of variable stars in $\omega$~Cen are summarized. Over
43 eclipsing binaries have been identified so far in the cluster field
of which only 8 objects are true detached systems.  Distance and age
determinations are presented for the system OGLEGC-17. The  apparent
distance modulus is $(m-M)_{V}=14.09\pm 0.04$ and the age of this
binary is $\tau=11.8\pm 0.6$~Gyr.
\end{abstract}

\section{Introduction}

Eclipsing binaries hold great promise as primary distance and age
indicators (Paczynski 1996). They can be used for establishing accurate
distances to globular clusters as well as to galaxies from the Local
Group.  The practical power of this method has been recently
demonstrated by Guinan et al. (1998) and Fitzpatrick et al. (2001) who
present the analysis of two binaries from the Large Magellanic Cloud.
That project is still in progress and data are being collected for
several systems.  A long term project known as DIRECT aims to determine
the distances to M31 and M33 through observations of eclipsing
binaries.  So far the main result obtained by DIRECT is the photometric
identification of a dozen or so systems suitable for detailed follow-up
observations with large telescopes (eg. Mochejska et al. 2001).
An extensive survey for eclipsing binaries in globular clusters
is  conducted by the CASE project (http://www.camk.edu.pl/case).

The binaries most suitable for the determination of ages of globular clusters
are detached systems hosting primaries which have recently finished
their evolution on the main-sequence. In such cases the age can be
calculated directly from the age-turnoff mass relation based on
stellar models.  The first practical application of this method was
presented recently by Thompson et al. (2001).

Attempts to use binaries to measure the distances and ages of globular
clusters have been hampered until recently by the paucity of suitable
systems. The situation is changing as several groups are undertaking
surveys for main sequence variables in globular clusters.  In
particular, the first detached eclipsing binaries were identified a few
years ago in the cluster $\omega$~Cen by the OGLE group (Kaluzny et al.
1996).  

In this contribution we summarize the results of searches for
eclipsing binaries in $\omega$~Cen. 
Subsequently, an analysis of the
eclipsing binary OGLEGC17 is presented.

\section{Photometric surveys for variable stars in $\omega$~Cen}

The first eclipsing binary in the field of $\omega$~Cen was discovered
by Martin (1938). Variable V78 has a period $P=1.17$~d and it is a
relatively bright object with $Vmax\approx 14.1$ (Sistero et
al. 1969). As it was shown first by Geyer \& Vogt (1978) this Algol
type variable is a foreground object unrelated to the cluster.\\
Subsequently Niss et al. (1978) searched for eclipsing binaries in the
central part of the cluster. Their survey was based on 11 photographic
plates collected over 5 nights with the newly dedicated 3.6-m
telescope at ESO. Seven candidates for eclipsing binaries were
identified.  One of them, star NJL5, turned out to be the first ever
eclipsing binary discovered in any globular cluster. NJL5 is a blue
straggler with a period $P=1.38$~d and an EB type light curve (Liller
1978; Jensen et al. 1985). It is a "radial velocity member" of the
cluster (Margon \& Cannon 1980).\\

An extensive survey for variable stars in $\omega$~Cen was undertaken
as a side project of the OGLE experiment (Kaluzny et al. 1996, 1997).
Six $15\times 15$ arcmin$^2$ fields were monitored with the 1.0-m Swope
telescope at Las Campanas Observatory. About 250 $V$-band frames were
collected for each field and 30 bona fide eclipsing binaries were
identified.  Most of these are contact systems but the sample includes
at least 3 detached binaries with EA-type light curves. Figure 1 shows
a schematic color magnitude-diagram of the cluster with the positions
of the eclipsing binaries identified by the OGLE team.  OGLEGC-14 with
$P=0.83$~d and $Vmax=16.64$ is a blue straggler composed of two A-type
stars.  Its $V$-band light curve is presented in Fig. 2.  The systemic
radial velocity of  OGLEGC-14 (unpublished) is compatible with cluster
membership, and this system deserves a detailed follow-up
investigation. OGLEGC-15 with $P=1.50$~d and $Vmax=16.96$ is located
slightly above the top of the cluster main-sequence on the
color-magnitude diagram. This binary seems to be composed of two
upper-main sequence stars. Light curves collected by the CASE group
during the last few years show significant season-to-season variability
suggesting that the system belongs to the RS CVn stars.  This
classification is supported by the recent identification of OGLEGC-15
as the optical counterpart of a relatively strong X-ray source (Cool
2002). The light curve of the system collected during the 2000 season
is presented in Fig. 3.

\begin{figure}
\plotfiddle{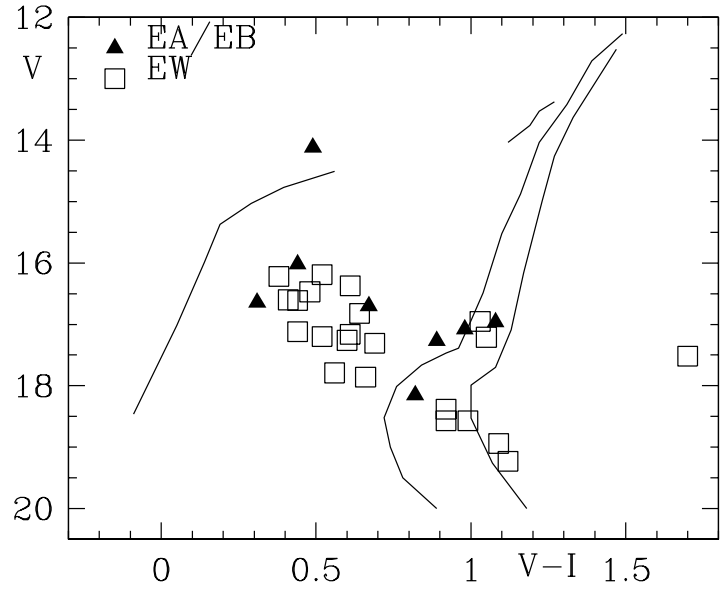}{8cm}{0}{70}{70}{-160}{-20}
\caption{The schematic CMD for $\omega$~Cen with the positions of 
eclipsing binaries identified by the OGLE survey marked.}
\end{figure}
%
\begin{figure}
\plotfiddle{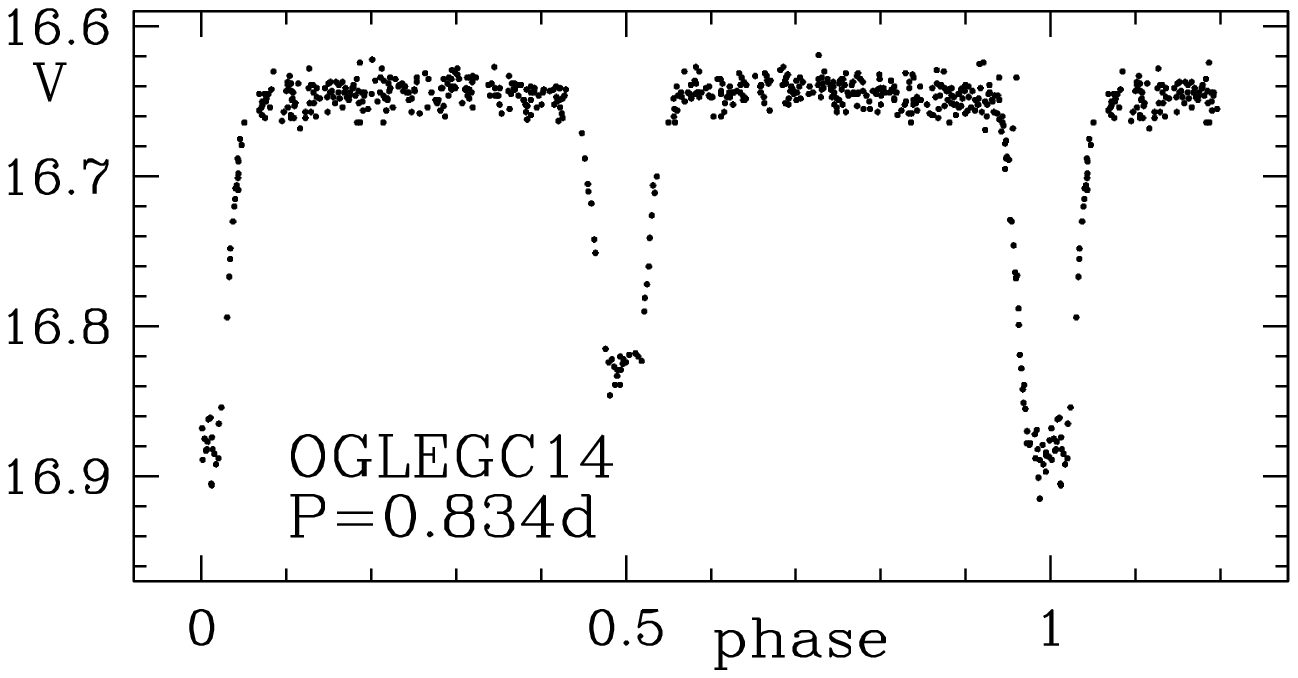}{7cm}{0}{100}{100}{-250}{-40}
\caption{The V-band light curve of OGLEGC-14 obtained by CASE.}
\end{figure}
\begin{figure}
\plotfiddle{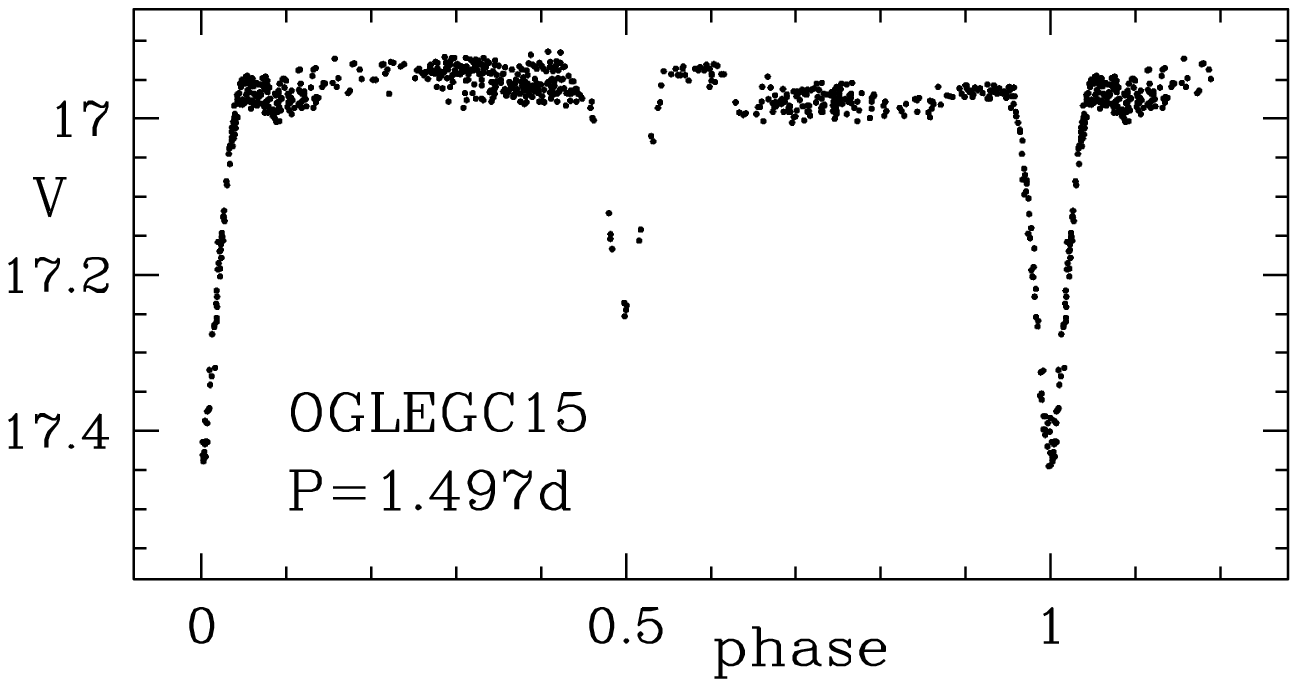}{7cm}{0}{100}{100}{-260}{-50}
\caption{The V-band light curve of OGLEGC-15 obtained by CASE.}
\end{figure}
The third detached EA-type system identified
by the OGLE team is OGLEGC-17. This binary was recently investigated
in detail by the CASE collaboration (Thompson et al. 2001).  The
analysis of spectroscopic and photometric data leads to the
determination of absolute parameters for both components of the
system.  In particular it was possible for the first time to measure
directly the masses of stars in any globular cluster. Moreover the
distance modulus to the cluster was derived using the surface
brightness method.  Below we present some new results on OGLEGC-17
which are based on  superior spectroscopy obtained recently with the
VLT UVES spectrograph.\\

During the 1999 and 2000 observing seasons the CASE group monitored
two $15\times 23$~arcmin$^2$ fields covering the main body of
$\omega$~Cen. The data were collected with the 1.0-m Swope telescope
and a total of 733 and 589 $V$-band frames were obtained for the east and
west fields, respectively. CCD images were analyzed with the ISIS
image-subtraction package (Alard 2000). 323 periodic variables were
identified: 165 RR~Lyr stars (15 new), 58 SX~Phe variables (32 new), and
42 eclipsing binaries (23 new). This sample includes 5 new bona fide
detached systems. A detailed description of these results will be
presented in a forthcoming paper.  (Kaluzny et al.; in preparation).
\section{Age and distance of OGLEGC-17}

Following the analysis presented by Thompson et al. (2001) we obtained
new radial velocity curves for OGLEGC-17. The new data came from the
VLT/UVES Echelle spectrograph. They consist of 22 spectra collected
near quadrature. From the combined spectra we measured the equivalent
widths for the CA II K-line near phases 0.25 and 0.75.  Knowing the
B-V color and the luminosity ratio of the components we were able to
derive a metallicity of $[Fe/H]=-2.29\pm 0.15$ from calibration of
Beers et al. (1990). This value places OGLEGC-17 in the low
metallicity tail of the distribution observed for the cluster. The
following masses of the components were derived:
$m1=0.809\pm0.012\odot$ and $m2=0.757\pm0.011\odot$. From the "age"
versus "turn-off mass" (see Thompson et al. 2001) we obtain
$\tau=11.80\pm0.6$~Gyr for the age of the primary component of
OGLEGC-17\footnote{
It is worth to note that this age determination is 
free from problems related to uncertainties of the "mixing length theory" 
of sub-photospheric convection. These uncertainties are potential source
of errors for methods relaying on comparison of observed and 
synthetic color-magnitude diagrams (eg. isochrone fitting).
}. 
Moreover, using the surface brightness method we derived
the apparent distance modulus of the binary $(m-M)_{V}=14.09\pm 0.04$.
A full analysis of the system based on the new radial velocity data will
be presented in a forthcoming paper (Thompson et al., in
preparation)


\begin{references}
\reference Alard, C. 2000, \aaps, 144, 363
\reference Beers, T.C.,  Kage, J.A., Preston, G.W., \& Shectman, S.A.
1990, \aj, 100, 849 
\reference Cool, A. 2002, in ASP Conf. Ser. Vol. ???, $\omega$~Centauri: A Unique
Window into Astrophysisc, ed.  F. van Leeuwen, G. Piotto \& J. Hughes 
(San Francisco, ASP), ???
\reference Fitzpatrik, E.L., Ribas, I., Guinan, E.F., 
DeWarf, L.E., Maloney, F.P., \& Massa, D., 2002, \apj, in press
\reference Geyer, E.H., \& Vogt, N. 1978, \aap, 67, 297
\reference Guinan, E.F. et al. 1998, \apj 509, L21
\reference Jensen, K.S, \& J{\o}rgensen, H.E. 1985, \aaps, 60, 229
\reference Kaluzny, J., Kubiak, M., Szymanski, M., Udalski, A., Krzeminski, W.,
\& Mateo, M. 1996, \aaps, 120, 139
\reference Kaluzny, J., Kubiak, M., Szymanski, M., Udalski, A., Krzeminski, W.,
\& Mateo, M. 1997, \aaps, 122, 471
\reference Margon, B., \& Cannon, R. 1989, Observatory, 109, 82
\reference Martin, W.C. 1938, Ann. van de Sterrenwachte Leiden, Deel XVII
\reference Liller, M.H. 1978, IBVS, No. 1527
\reference Mochejska, B.J., Kaluzny, J., Stanek, K.Z.,
 Sasselov, D. D.,\& Szentgyorgyi, A.H. 2001, \aj, 121, 2032
\reference Niss, B., J{\o}rgensen, H.E., \& Lautsen, S. 1978, \aaps, 32, 387
\reference Paczy\'nski, B. 1996, in Space Telescope Science Institute Series, 
The Extragalactic  Distance Scale, ed. M. Livio, 
(Cambridge; Cambridge Univ. Press), 273
\reference Sistero, R.F., Fourcade, C.R., \& Laborde, J.R. 1969, IBVS,  No. 402
\reference Thompson, I.B. et al. 2001, \aj, 121, 3089  

\end{references}
\end{document}